\documentstyle[12pt]{article}

\font\twlgot =eufm10 scaled \magstep1
\font\egtgot =eufm8
\font\sevgot =eufm7

\font\twlmsb =msbm10 scaled \magstep1
\font\egtmsb =msbm8
\font\sevmsb =msbm7

\newfam\gotfam

\textfont\gotfam\twlgot
\scriptfont\gotfam\egtgot
\scriptscriptfont\gotfam\sevgot

\newfam\msbfam
\textfont\msbfam\twlmsb
\scriptfont\msbfam\egtmsb
\scriptscriptfont\msbfam\sevmsb
\def\Bbb{\protect\pBbb}
\def\pBbb{\relax\ifmmode\expandafter\Bb\else\typeout{You cann't use
Bbb in text mode}\fi}
\def\Bb #1{{\fam\msbfam\relax#1}}

\def\op#1{\mathop{{\it\fam0} #1}\limits}

\newcommand{\lng}{\langle}
\newcommand{\rng}{\rangle}

\newcommand{\bite}{\begin{itemize}}
\newcommand{\eite}{\end{itemize}}
\newcommand{\benu}{\begin{enumerate}}
\newcommand{\eenu}{\end{enumerate}}
\newcommand{\bde}{\begin{description}}
\newcommand{\ede}{\end{description}}
\newcommand{\bquo}{\begin{quote}}
\newcommand{\equo}{\end{quote}}
\newcommand{\bquot}{\begin{quotation}}
\newcommand{\equot}{\end{quotation}}

\newcommand{\eqref}[1]{(\ref{#1})}
\newcommand{\beq}{\begin{equation}}
\newcommand{\eeq}{\end{equation}}
\newcommand{\ben}{\begin{eqnarray}}
\newcommand{\een}{\end{eqnarray}}
\newcommand{\be}{\begin{eqnarray*}}
\newcommand{\ee}{\end{eqnarray*}}
\newcommand{\bea}{\begin{eqalph}}
\newcommand{\eea}{\end{eqalph}}

\newcommand{\bb}{{\bf 1}}

\newcommand{\al}{\alpha}

\newcommand{\dl}{\delta}
\newcommand{\la}{\lambda}

\newcommand{\f}{\phi}

\newcommand{\m}{\mu}

\newcommand{\e}{\epsilon}

\newcommand{\th}{\theta}

\newcommand{\wt}{\widetilde}

\newcommand{\ol}{\overline}

\newcommand{\dr}{\partial}

\newcommand{\ot}{\otimes}
\newcommand{\ap}{\approx}

\newcounter{theorem}
\newcounter{remark}
\newcounter{proposition}
\newcounter{lemma}
\newcounter{corollary}
\newcounter{definition}

\def\theremark{\arabic{remark}}

\def\thedefinition{\arabic{definition}}

\newcommand{\mar}[1]{}

\begin{document}
\hbox{}

{\parindent=0pt

{\large\bf  Hopf algebras of canonical
commutation relations}
\bigskip

{\sc G. Sardanashvily}
\medskip

\begin{small}

Department of Theoretical Physics,
Physics Faculty, Moscow State
University, 117234 Moscow, Russia

E-mail: sard@grav.phys.msu.su

URL: http://webcenter.ru/$\sim$sardan/
\bigskip

{\bf Abstract}

Given a Heisenberg algebra $A$ of canonical commutation relations
modelled over an infinite-dimensional nuclear space, a Hopf algebra
of its quantum deformations is also an algebra of canonical commutation
relations whose Fock representation recovers some non-Fock
representation of $A$.

\end{small}
}

\section{Introduction}

By virtue of the well-known Stone--von Neumann uniqueness theorem,
all irreducible representations of the
canonical commutation relations (henceforth the CCR)
of finite degree of freedom are equivalent.
On the contrary, the infinite-dimensional CCR possess many
non-equivalent irreducible representations (see \cite{flor}
for a survey). 
Here, we restrict our consideration to the CCR modelled over
a nuclear space. They include the CCR of
finite degrees of freedom, but we focus on the infinite-dimensional CCR.
In particular,
this is the case
of field theory \cite{sard02}.

Let $A$ be the Heisenberg algebra of the CCR
modelled over a nuclear space. 
Since $A$ is a Lie algebra,
one can associate to $A$ a Hopf algebra, regarded as an algebra of
$q$-deformed CCR (see \cite{ior} for the case of finite-dimensional CCR). 
We show that this Hopf algebra is the enveloping algebra of another CCR
algebra $A_{q,c}$. Moreover, $A$ and
$A_{q,c}$ possess the same set of representations. Herewith, operators
of the Fock
representation of $A_{q,c}$ carry out some non-Fock representation 
of $A$. 

\section{The nuclear CCR}

Let us recall the notion of a nuclear space (see, e.g., \cite{piet}).
Let a complex vector space $V$ be provided with a countable set of
non-degenerate
Hermitian forms $\lng.|.\rng_k$, $=1,\ldots$, such that
\be
\lng v|v\rng_1\leq \cdots\leq \lng v|v\rng_k\leq\cdots
\ee
for all $v\in V$. Let $V$ be complete in the topology defined by the
set of norms $\|.\|^{1/2}_k=\lng .|.\rng_k$.
Then $V$ is called a countably Hilbert space.
Let $V_k$ denote the completion of $V$ with respect to the norm
$\|.\|_k$. There is the chain of injections
$V_1\supset V_2\supset \cdots V_k\supset \cdots$,
and $V=\op\cap_k V_k$.
Let $T^n_m$, $m\leq n$, be a
prolongation of the map
$V_n\supset V\ni v\mapsto v\in V\subset V_m$
to the continuous map of $V_n$ onto a dense subset of $V_m$. A
countably Hilbert space $V$
is called a nuclear space if, for any $m$,
there exists $n$ such that $T^m_n$ is a nuclear map, i.e.,
\be
T^n_m(v)=\op\sum_i\la_i\lng v|v^i_n\rng_{V_n} v^i_m,
\ee
where: (i)  $\{v^i_n\}$ and $\{v_m^i\}$ are bases for the
Hilbert spaces $V_n$ and $V_m$, respectively, (ii) $\la_i\geq 0$, and
(iii) the series
$\sum \la_i$ converges.
Note that a Hilbert
space is not nuclear, unless it
is finite-dimensional. 

Let $V$ be a real nuclear space provided with still another non-degenerate
Hermitian form $\lng.|.\rng$, which is separately continuous. This form
makes $V$ to a separable pre-Hilbert space.
Let us consider the
group $G$
of the triples $(v_1,v_2,\la)$ of elements $v_1$, $v_2$
of $V$ and complex numbers $\la$ of unit modulus which are subject to
multiplications
\mar{qm541}\beq
(v_1,v_2,\la)(v'_1,v'_2,\la')=(v_1+v'_1,v_2+v'_2, \exp[i\lng
v_2,v'_1\rng] \la\la'). \label{qm541}
\eeq
It is a Lie group whose
group space is a nuclear manifold modelled over 
\mar{h1}\beq
W=V\oplus V\oplus \Bbb R. \label{h1}
\eeq
Let us denote $T(v)=(v,0,0)$ and $P(v)=(0,v,0)$.
Then the multiplication law (\ref{qm541}) takes the form
\mar{qm543}\ben
&& T(v)T(v')=T(v+v'),\qquad P(v)P(v')=P(v+v'),\nonumber\\
&& P(v)T(v')=\exp[i\lng v|v'\rng]T(v')P(v). \label{qm543}
\een
Written in this form, $G$ is called the Weyl CCR group.

The Lie algebra of the nuclear Lie group $G$ is the
above mentioned Heisenberg algebra $A$. It is
generated by the Hermitian elements $I$,
$\f(v)$, $\pi(v)$, $v\in V$,
which obey the commutation relations
\mar{qm540,'}\ben
&& [\f(v),I]=[\pi(v),I]=[\f(v),\f(v')]=[\pi(v),\pi(v')]=0, \label{qm540}\\
&&  [\pi(v),\f(v')]=-i\lng v|v'\rng I. \label{qm540'}
\een
Given a countable orthonormal basis $\{v_i\}$ for the pre-Hilbert space $V$,
the CCR (\ref{qm540}) -- (\ref{qm540'}) take the form
\be
[\f(v_j),\f(v_k)]=[\pi(v_k),\pi(v_j)]=0, \qquad
[\pi(v_j),\f(v_k)]=-i\dl_{jk}I.
\ee

One also introduces the creation and
annihilation operators
\mar{qm552}\beq
a^\pm(v)=\frac{1}{\sqrt 2}[\f(v)\mp i\pi(v)]. \label{qm552}
\eeq
They obey the conjugation rule $(a^\pm(v))^*=a^\mp(v)$
and the commutation relations
\be
[a^-(v), a^+(v')]=\lng v|v'\rng I, \qquad
[a^+(v),a^+(v')]=[a^-(v),a^-(v')]=0.
\ee

\section{Hopf algebras of the CCR}

Let us consider the tensor algebra $\ot W$ of the vector space $W$
(\ref{h1}) generated by elements $\f(v)$, $\pi(v)$ and $I$. 
It is provided with a unique Hopf algebra structure,
characterized by the comultiplication
\be
\Delta(w)=w\ot\bb +\bb\ot w, \qquad w\in W,
\ee
the counit $\e(w)=0$, the antipode $S(w)=-w$, and the universal
matrix $R=\bb\ot\bb$. It is a cocommutative quasi-triangular Hopf
algebra, called the classical Hopf algebra.

Let $\ol A$ be the enveloping algebra of the Heisenberg CCR algebra $A$.
It is the quotient of the tensor algebra $\ot W$ by the commutation
relations (\ref{qm540}) -- (\ref{qm540'}), written with respect to the
tensor product $\ot$, and by the relation 
\mar{h2}\beq
I\ot I=I. \label{h2}
\eeq
The $\ol A$ inherits the structure of the classical Hopf algebra on
$\ot W$. 
We denote it $B_{\rm cl}(A)$.

Now let us consider the quotient $\ol A_{q,c}$ of the 
tensor algebra $\ot W$ by the
relations (\ref{qm540}), (\ref{h2}) and the commutation relations
\mar{h3}\beq
[\pi(v),\f(v')]=-i\lng v|v'\rng\frac{q^{cI}-q^{-cI}}{c(q-q^{-1})}, 
\label{h3}
\eeq
where $q$ and $c$ are strictly positive real numbers. Due to the relation
(\ref{h2}), the right-hand side of the
relations (\ref{h3}) is well defined on $\ot W$, and we have
\mar{h4}\beq
[\pi(v),\f(v')]=-i\lng v|v'\rng\frac{q^c-q^{-c}}{c(q-q^{-1})}I.
\label{h4}
\eeq
Hence, $\ol A_{q,c}$ is the enveloping algebra of the Heisenberg CCR algebra
$A_{q,c}$ given by the commutation relations (\ref{qm540}) and (\ref{h4}).
This CCR algebra is modelled over the same nuclear space $V$, but
provided with the Hermitian form 
\mar{h5}\beq
\lng v|v'\rng_{q,c}=C_{q,c}\lng v|v'\rng, \qquad
C_{q,c}=\frac{q^c-q^{-c}}{c(q-q^{-1})}. \label{h5}
\eeq
The enveloping algebra $\ol A_{q,c}$ admits both the structure of the
classical Hopf algebra $B_{\rm cl}(A_{q,c})$ and the Hopf algebra 
$B(A_{q,c})$, which differs from the
classical one in the comultiplication law
\be
&& \Delta(\f(v))=\f(v)\ot q^{cI/2} + q^{-cI/2}\ot \f(v), \qquad
\Delta(\pi(v))=\pi(v)\ot q^{cI/2} + q^{-cI/2}\ot \pi(v),\\
&& \Delta(I)=I\ot\bb +\bb\ot I. 
\ee
One can think of $B(A_{q,c})$ as being a Hopf algebra of the $q$-deformed
CCR.
It is readily observe that, if $c=1$, the CCR algebras $A$ and
$A_{q,1}$ coincide for any $q$, but the Hopf algebra $B(A_{q,1})$
differs from the classical one $B_{\rm cl}(A_{q,1})=B_{\rm cl}(A)$. If $q=1$,
then $A_{1,c}=A$ and $B(A_{1,c})=B_{\rm cl}(A)$ for any $c$.

Since the Hopf algebra $B(A_{q,c})$ is the enveloping algebra of the
CCR algebra $A_{q,c}$, its representations are determined in full by
representations of $A_{q,c}$
Let us compare the representations of the CCR algebras $A$ and $A_{q,c}$.

\section{Representations of the nuclear CCR}

The CCR group $G$
contains two  Abelian subgroups $T$ and $P$.
Following the
representation algorithm in \cite{gelf64},
we first construct representations
of the nuclear Abelian group $T$ \cite{sard02}.

Its cyclic strongly continuous unitary representation $\rho$
in a Hilbert space
$(E,\lng.|.\rng_E)$ with a (normed) cyclic vector $\theta\in E$ defines
the complex function
\be
Z(v)=\lng \rho(T(v))\theta|\theta\rng_E
\ee
on $V$. This function is 
continuous and positive-definite, i.e., $Z(0)=1$ and
\be
\op\sum_{i,j} Z(v_i-v_j)\ol c_i c_j\geq 0
\ee
for any finite set $v_1,\ldots,v_m$ of elements of $V$ and arbitrary
complex numbers $c_1,\ldots,c_m$.
By virtue of the well-known Bochner theorem,
such a function on a nuclear space $V$ is the Fourier
transform
\mar{qm545}\beq
Z(v)=\int\exp[i\lng v,u\rng]\m \label{qm545}
\eeq
of a positive measure $\m$ of total mass 1 on the topological dual $V'$ of
$V$. Then the above mentioned
representation $\rho$ of $T$ can be given by the operators
\mar{q2}\beq
T_Z(v)f(u)=\exp[i\langle v,u\rangle]f(u)  \label{q2}
\eeq
in the Hilbert space $L^2(V',\m)$ of classes of $\m$-equivalent
square integrable complex
functions $f(u)$ on $V'$.
The cyclic vector $\th$ of this representation is the
$\m$-equivalence class $\th\ap_\m 1$ of the constant function $f(u)=1$.
Conversely, every positive measure $\m$ of total mass 1 on the dual $V'$
of $V$ (and, consequently, every continuous positive-definite
function $Z(v)$ on $V$) defines a cyclic strongly continuous unitary
representation (\ref{q2}) of
the nuclear group $T$. 
We agree to call $Z$ a
generating function of this representation.
One can show that distinct generating functions $Z$ and $Z'$ determine
equivalent representations $T_Z$ and $T_{Z'}$ (\ref{q2}) of $T$ in
the Hilbert spaces $L^2(V',\m)$ and $L^2(V',\m')$  iff they are
the Fourier transform of equivalent measures on $V'$. 

The representation $T_Z$ (\ref{q2}) of the  group
$T$ can be extended to
the CCR group $G$ if the measure $\m$ possesses the following property.
Let $u_v$, $v\in V$, denote an element of $V'$ given by the condition
\mar{qm546}\beq
\lng v',u_v\rng=\lng v'|v\rng, \qquad \forall v'\in V. \label{qm546}
\eeq
These elements form the image of the monomorphism
$V\to V'$ determined by the Hermitian form $\lng.|.\rng$ on $V$.
Let the measure $\m$ in (\ref{qm545}) remain equivalent under translations
$u\mapsto u+u_v$ of $V'$ by any element $u_v$ of $V\subset V'$,
i.e.,
\mar{qm547}\beq
\m(u+u_v)=a^2(v,u)\m(u), \qquad \forall u_v\in V\subset V', \label{qm547}
\eeq
where a function $a(v,u)$ is square $\m$-integrable and strictly
positive almost everywhere on $V'$. This function fulfils the relations
\mar{qm555}\beq
a(0,u)=1, \qquad a(v+v',u)=a(v,u)a(v',u+u_v). \label{qm555}
\eeq
A measure on $V'$ obeying the condition (\ref{qm547}) is called
translationally quasi-invariant.
Let the generating function $Z$ of a cyclic strongly continuous unitary
representation of the nuclear group $T$ be the Fourier transform
(\ref{qm545}) of such a measure $\m$ on $V'$. 
Then the representation
(\ref{q2}) of $T$ is extended to the representation of the nuclear CCR
group $G$ in the Hilbert space $L^2(V',\m)$ by operators
\mar{qm548}\beq
P_Z(v)f(u)=a(v,u)f(u+u_v). \label{qm548}
\eeq
Moreover, one can show that if
$\m'$ is a $\m$-equivalent positive measure
of total mass 1 on $V'$, it is also translationally quasi-invariant
and provides an equaivalent representation of $G$.

A strongly continuous
unitary representation $T_Z$ (\ref{q2}), $P_Z$ (\ref{qm548}) of the
nuclear CCR group $G$ implies a
representation
of its Lie algebra $A$ by (unbounded) operators  
\mar{qm549,62}\ben
&& I=\bb,\quad \f(v)f(u)=\lng v,u\rng f(u), \quad
   \pi(v) f(u)=-i(\dl_v+\eta(v,u))f(u), \label{qm549}\\
&& \dl_v f(u)=\op\lim_{\al\to 0}\al^{-1}[f(u+\al u_v)-f(u)],
\qquad \al\in\Bbb R,\nonumber\\
&& \eta(v,u)=\op\lim_{\al\to 0}\al^{-1}[a(\al v,u)-1],\label{qm562}
\een
in the same
Hilbert space $L^2(V',\m)$.
With the aid of the formulas 
\be
&& \dl_v\dl_{v'}=\dl_{v'}\dl_v, \qquad
\dl_v(\eta(v',u))=\dl_{v'}(\eta(v,u)),\\
&& \dl_v=-\dl_{-v}, \qquad \dl_v(\lng v',u\rng)=\lng v'|v\rng,\\
&& \eta(0,u)=0, \quad \forall u\in V',\qquad
\dl_v\th=0,  \quad \forall v\in V,
\ee
derived from the relations (\ref{qm555}),
it is easily justified that the operators
(\ref{qm549}) fulfil the Heisenberg CCR (\ref{qm540}).

Gaussian measures exemplify a physically relevant class of
translationally quasi-invariant measures on the dual $V'$ of a nuclear
space $V$. The Fourier transform of a Gaussian measure reads
\mar{spr523}\beq
Z(v)=\exp\left[-\frac12 M(v)\right], \label{spr523}
\eeq
where $M(v)$ is a seminorm on $V'$ called the covariance form.
Let $\m_K$ denote a Gaussian measure on $V'$
whose Fourier transform is the generating function
\mar{qm563}\beq
Z_K=\exp[-\frac12 M_K(v)] \label{qm563}
\eeq
with the
covariance form $M_K(v)=\lng K^{-1}v|K^{-1}v\rng$, 
where $K$ is a bounded invertible operator in the Hilbert completion $\wt V$
of $V$ with respect to the Hermitian form $\lng.|.\rng$.
The Gaussian measure
$\m_K$ is translationally quasi-invariant:
\mar{qm561}\ben
&& \m_K(u+u_v)=a_K^2(v,u)\m_K(u), \nonumber \\
&&a_K(v,u)= \exp[-\frac14 M_K(Cv)-
\frac12\lng Cq,u\rng], \label{qm561}
\een
where $C=KK^*$ is a bounded Hermitian operator in $\wt V$.

Let us construct the representation of the CCR algebra $A$
determined by the generating function $Z_K$ (\ref{qm563}).
Substituting the function (\ref{qm561})
into the formula (\ref{qm562}), we find
\be
\eta(v,u)= -\frac12\lng Cv,u\rng.
\ee
Hence, the operators $\f(v)$ and $\pi(v)$ (\ref{qm549}) take the form
\mar{qm565}\beq
\f(v)=\lng v,u\rng, \qquad
\pi(v)=-i(\dl_v-\frac12\lng Cv,u\rng). \label{qm565}
\eeq
Accordingly, the creation and annihilation operators (\ref{qm552})
read
\mar{qm566}\beq
a^\pm(v)=\frac{1}{\sqrt 2}[\mp\dl_v \pm \frac12\lng Cv,u\rng + \lng
v,u\rng]. \label{qm566}
\eeq

In particular, let us put $K=\sqrt2\cdot\bb$.
Then the generating function (\ref{qm563}) takes the form
\mar{qm567}\beq
Z_{\rm F}(v)=\exp[-\frac14\lng v|v\rng], \label{qm567}
\eeq
and determines the Fock
representation
of the CCR algebra $A$ by the operators
\be
&& \f(v)=\lng v,u\rng, \qquad  \pi(v)=-i(\dl_v-\lng v,u\rng), \\
&& a^+(v)=\frac{1}{\sqrt 2}[-\dl_v + 2\lng v,u\rng], \qquad
a^-(v)=\frac{1}{\sqrt 2}\dl_v.
\ee

Note that the Fock representation up to an equivalence is characterized by
the existence of a cyclic vector
$\th$ such that
\mar{qm570}\beq
a^-(v)\th=0, \qquad \forall v\in V. \label{qm570}
\eeq
An equivalent condition is that there exists the particle number
operator $N$ possessing a lower bounded spectrum.
This operator is defined 
by the conditions
\be
[N,a^\pm(v)]=\pm a^\pm(v)
\ee
up to a summand $\la\bb$. With respect to a countable orthonormal
basis $\{v_k\}$, it is given by the sum
\be
N=\op\sum_k a^+(v_k)a^-(v_k). 
\ee
A glance at the expression (\ref{qm566}) shows that the condition
(\ref{qm570}) does not hold, unless $Z_K$ is $Z_{\rm F}$ (\ref{qm567}).
For instance, the particle number operator in the representation
(\ref{qm566}) reads
\be
&& N=\op\sum_j a^+(v_j)a^-(v_j)= \op\sum_j[-\dl_{v_j}\dl_{v_j}
+C^j_k\lng v_k,u\rng\dr_{v_j} + \\
&& \qquad
(\dl_{km}-\frac14 C^j_kC^j_m)\lng v_k,u\rng\lng v_m,u\rng -
(\dl_{jj}-\frac12 C^j_j)].
\ee
One can show that this operator is
defined and is lower bounded  
only if the operator $C$
is a sum of the scalar operator $2\cdot\bb$ and a nuclear operator in
$\wt V$. For instance, the generating function
\be
Z_c(v)=\exp[-\frac{c^2}{2}\lng v|v\rng], \qquad c^2\neq \frac12,
\ee
determines a non-Fock representation
of the nuclear CCR.

At the same time, the non-Fock representation (\ref{qm565}) of the
CCR algebra (\ref{qm540}) is the Fock representation
\be
&& \f_K(v)=\f(v)=\lng v,u\rng,\\
&& \pi_K(v)=\pi(S^{-1}v) = -i(\dl^K_v -\frac12\lng v,u\rng), \qquad
\dl^K_v =\dl_{S^{-1}v},
\ee
of the CCR algebra $\{\f_K(v),\pi_K(v),I\}$, where
\be
[\f_K(v),\pi_K(v)]=i\lng K^{-1}v|K^{-1}v'\rng I.
\ee

Bearing in mind this fact, turn now to the CCR algebra $A_{q,c}$ in Section 3. 
Comparing the commutation relations (\ref{qm540'}) and (\ref{h4}), 
one can show that, given a representation $\rho$ of the CCR algebra
$A$, the CCR algebra $A_{q,c}$ admits a representation $\rho_{q,c}$ 
by the operators
\be
\rho_{q,c}(\f(v))=\rho(\f(v)), \qquad
\rho_{q,c}(\pi(v))=\rho(\pi(C_{q,c}v)), \qquad 
\rho_{q,c}(I)=\rho(I)=\bb,
\ee
where $C_{q,c}$ is the real number given by the expression (\ref{h5}).
For instance, if $\rho$ is the Fock representation of the CCR algebra 
$A$, the representation $\rho_{q,c}$ is not equivalent to the Fock
representation of the CCR algebra $A_{q,c}$, unless $V$ is
finite-dimensional.

\end{document}